\documentclass[pdftex,twocolumn,epjc3,preprintnumbers,amsmath,amssymb]{svjour3}          

\RequirePackage[T1]{fontenc}

\smartqed  

\RequirePackage{graphicx}
\RequirePackage{mathptmx}   
\RequirePackage{flushend}
\RequirePackage[numbers,sort&compress]{natbib}
\RequirePackage[colorlinks,citecolor=blue,urlcolor=blue,linkcolor=blue]{hyperref}

\journalname{Eur. Phys. J. C}
\usepackage{amsmath}
\usepackage{amssymb}
\usepackage[normalem]{ulem}
\usepackage{bm}
\usepackage{color}
\usepackage{xcolor}
\usepackage{soul}
\usepackage{graphicx}
\usepackage[colorlinks,citecolor=blue,urlcolor=blue,linkcolor=blue]{hyperref}
\usepackage{multirow}
\usepackage{subfigure}
\usepackage{booktabs}
\usepackage{mathrsfs}
\usepackage{balance}
\setulcolor{yellow}

\newcommand{\f}{\frac}
\newcommand{\lt}{\left}

\newcommand{\rt}{\right}

\newcommand{\dt}{\delta}

\begin{document}
\title{Exploring sub-GeV Dark Matter Physics with Cosmic Ray and Future Telescopes} 

\author{Guan-Sen Wang\thanksref{addr1,addr2} 
\and
Bing-Yu Su\thanksref{addr1,addr2,e1} 
\and 
Lei Zu\thanksref{addr3,e2}
\and 
Lei Feng\thanksref{addr1,addr2,e4}} 

\thankstext{e1}{e-mail: bysu@pmo.ac.cn (corresponding author)}
\thankstext{e2}{e-mail: lei.zu@astro.nao.ac.jp (corresponding author)}
\thankstext{e4}{e-mail: fenglei@pmo.ac.cn (corresponding author)}

\institute{Key Laboratory of Dark Matter and Space Astronomy, Purple Mountain Observatory, Chinese Academy of Sciences, Nanjing 210023, China\label{addr1} 
\and 
School of Astronomy and Space Science, University of Science and Technology of China, Hefei, Anhui 230026, China\label{addr2} 
\and 
National Centre for Nuclear Research, ul. Pasteura 7, 02-093 Warsaw, Poland\label{addr3} 
}

\date{Received: date / Accepted: date}

\maketitle

\begin{abstract}
If sub-GeV dark matter (DM) annihilates into charged particles such as $e^+ e^-$, $\mu^+ \mu^-$, or $\pi^+ \pi^-$, it generates an additional source of electrons and positrons in the cosmic ray (CR) population within our Milky Way. During propagation, these secondary electrons and positrons undergo reacceleration processes, boosting their energies to the GeV scale. Observatories like AMS-02 can detect these high-energy particles, enabling constraints on the properties of sub-GeV DM. By analyzing AMS-02 electron and positron data, the 95\% upper limits on the DM annihilation cross-section have been established in the range of $10^{-28}$ to $10^{-27}$ cm$^3\,$s$^{-1}$, corresponding to DM masses ranging from 100 MeV to 1 GeV. Meanwhile, MeV telescopes will provide complementary constraints on DM properties by detecting photon emissions from such annihilation processes. Notably, the sensitivity of future MeV gamma-ray observatories is projected to approach or match the constraints derived from CR data.

\end{abstract}

\section{Introduction}
Over the past decades, dark matter (DM) has remained one of the most pressing mysteries in astrophysics and cosmology~\cite{Cirelli:2024ssz}. While extensive evidence for its existence has been gathered from galactic rotation curves~\cite{rotationcurves,Jiao:2023aci,Ou:2023adg,Labini:2023fmy}, gravitational lensing~\cite{Clowe:2006eq,vanWaerbeke:2000rm,Massey:2007wb,Harvey:2015hha,Robertson:2016qef}, and observations of the cosmic microwave background (CMB)~\cite{Lewis:2006fu,Hanson:2009kr,Planck:2018lbu},
the precise nature and interaction mechanisms of DM remain elusive. Historically, much of the research has focused on DM candidates in the GeV--TeV mass range, commonly referred to as weakly interacting massive particles (WIMPs). These particles naturally explain the observed DM relic density through the thermal freeze-out mechanism \cite{Feng:2013zca, Pospelov:2007mp, Alekhin:2015byh, Roszkowski:2017nbc, Slatyer:2009yq, XENON:2020kmp, Gaskins:2016cha}. Despite decades of dedicated effort, no conclusive evidence for WIMPs has been found. A wide range of experiments designed to probe WIMP interactions has yielded null results. Direct detection experiments such as XENONnT, LUX, and PandaX \cite{XENON:2023cxc,XENON:2018voc,LUX:2016ggv,LUX:2013afz,PandaX-II:2017hlx,PandaX-II:2016vec} have excluded significant portions of the parameter space for DM-baryon and DM-electron interactions for WIMPs with masses above the GeV scale. Similarly, indirect searches through cosmic ray (CR) and $\gamma$-ray observations such as AMS-02~\cite{Abdughani:2021pdc,Cui:2016ppb,Giesen:2015ufa,Cholis:2019ejx,Cuoco:2016eej,Nguyen:2024kwy} and Fermi-LAT~\cite{Baring:2015sza,Fermi-LAT:2016uux,MAGIC:2016xys,Fermi-LAT:2013thd,Muru:2025vpz} have placed strong constraints on DM annihilation. These results have shifted the attention of researchers toward exploring alternative DM mass ranges, particularly in the MeV (sub-GeV) scale~\cite{Cirelli:2025rky,Horoho:2025knb,Cheek:2025nul,Wang:2023sxr, Su:2024hrp, Peter:2012rz, Bertone:2016nfn, Arun:2017uaw, Boudaud:2016mos, Zu:2021odn, Cirelli:2020bpc, Boddy:2015efa, Essig:2013goa, Wadekar:2021qae, Caputo:2022dkz, Cirelli:2023tnx, DelaTorreLuque:2023olp, ODonnell:2024aaw, Saha:2025wgg,Balaji:2025afr}. Such low-mass DM candidates present new opportunities and challenges for detection, potentially offering critical insights into the nature of DM.


However, both traditional direct and indirect detection methods for DM face significant challenges when probing the sub-GeV mass range. On the one hand, direct detection relies on nuclear recoil, which loses sensitivity when DM mass falls far below GeV-scale nuclei. This suppresses sub-GeV DM signals. On the other hand, indirect detection methods also face challenges in this mass range due to the absence of sensitive instruments. 
In particular, MeV astrophysics, lying between the well-explored keV scale of X-ray astronomy and the GeV--TeV range of $\gamma$-ray astrophysics, has historically suffered from a lack of observations and understanding. This is often referred to as the ``MeV gap''~\cite{Carenza:2022som,Boddy:2015fsa}.
To overcome this gap, several groups have developed methods that exploit the X-ray signal arising from secondary inverse-Compton emission produced by DM-induced electrons~\cite{Cirelli:2023tnx,Cirelli:2020bpc,Balaji:2025afr}. Conversely, other groups have considered CR reacceleration mechanisms~\cite{Boudaud:2016mos,Zu:2021odn}, which boost sub-GeV electrons and positrons to the GeV scale, making them detectable with current CR observatories such as AMS-02.

Actually, beyond reacceleration effect, convective effect caused by galactic winds is also commonly considered. However, analysis of precise nuclei data (including AMS-02 B/C ratios, proton fluxes, and primary/secondary spectra like C, O, B, alongside ACE and Voyager measurements) consistently shows that only propagation models with reacceleration can fit both primary cosmic ray fluxes and secondary-to-primary ratios~\cite{Yuan:2017ozr,Yuan:2018vgk,Yuan:2018lmc}. Furthermore, the reacceleration effect is key to naturally explaining the observed spectral hardening, particularly the fact that secondary nuclei spectra harden more than primary spectra at high energies~\cite{Yuan:2018lmc}. This occurs because reacceleration causes a steepening of secondary-to-primary ratios at low energies, allowing the differing hardening patterns to be explained by source injection spectra alone, without requiring additional propagation-related spectral breaks. While reacceleration is thus essential for modeling CR propagation, and has been applied in previous works like Ref. \cite{Boudaud:2016mos} to constrain sub-GeV DM, such analyses yielded relatively crude due to the complex CR Backgrounds and uncertain propagation parameters. However, recent advances in detailed CR backgrounds~\cite{Yuan:2013eja,Yuan:2013eba} and propagation~\cite{Yuan:2017ozr,Yuan:2018vgk,Yuan:2018lmc} now enable us to achieve significantly improved results.

Thus, in this work, we utilize the CR reacceleration effect to calculate the propagation of secondary electrons and positrons produced from DM annihilation. Following previous research, we explore additional annihilation channels and account for uncertainties in CR propagation parameters. Specifically, we consider channels such as $e^+e^-$, $\mu^+\mu^-$, 4 electrons, 4 muons and $\pi^+\pi^-$ channels. We also investigate the potential impact of uncertain CR propagation parameters. By analyzing AMS-02 electron and positron data, we derive constraints on the DM annihilation cross-section, which range from $10^{-28}$ to $10^{-27}$ cm$^3$ s$^{-1}$, corresponding to the DM mass from 100 MeV to 1 GeV. These bounds are comparable to those obtained from CMB data and are significantly more stringent than limits derived from Voyager observations~\cite{Boudaud:2016mos,Stone:2013zlg}. 

Additionally, we calculate the MeV photon flux resulting from final state radiation or the decay of annihilation products. Under optimistic estimations, DM-rich regions are capable of producing sufficient MeV photons to be detectable by upcoming MeV gamma-ray telescopes, such as AMEGO~\cite{AMEGO:2019gny}, AMEGO-X~\cite{Fleischhack:2021mhc}, MEGA~\cite{And:2005}, GRIPS~\cite{Greiner:2008yd}, e-Astrogam~\cite{e-ASTROGAM:2017pxr} and Very Large Area gamma-ray Space Telescope (VLAST)~\cite{fanyizhong}. This indicates that the high sensitivity MeV telescope in the future are crucial for exploring the physics of sub-GeV DM. The detection of MeV photons complements CR observations, as it provides an independent and direct probe of DM annihilation or decay processes. While CR experiments like AMS-02 are sensitive to the charged particle products of DM interactions, MeV gamma-ray telescopes can capture the secondary electromagnetic signatures, offering a more comprehensive picture of DM behavior. 

This work is organized as follows: In Sec.~\ref{sec:ep}, we analyse the electrons and positrons spectrum from DM annihilation, and present the properties of the Milky Way. Then we calculate the background CR spectrum in Sec.~\ref{sec:back}, followed by a summary of the results related to CRs in Sec.~\ref{sec:res}. Additionally, in Sec.~\ref{sec:gamma}, we calculate the gamma-ray radiation produced by DM annihilation. Finally, we conclude in Sec.~\ref{sec:con}.

\section{Modeling Sub-GeV DM Annihilation and CR Propagation} \label{sec:ep}

In this work, we investigate sub-GeV DM annihilation in the Milky Way by utilizing AMS-02 electron and positron data, incorporating the effects of reacceleration. To describe the DM density profile in our Milky Way, we use NFW profile~\cite{Navarro:1996gj} 
\begin{eqnarray}
\rho_{\rm NFW}=\frac{\rho_{\rm s}}{r/r_{\rm s}(1+r/r_{\rm s})^2}.
\label{eqnfw}
\end{eqnarray}
Moreover, due to the limitations of current observations and the uncertainties in modeling, the specific distribution of DM within the Milky Way remains subject to considerable uncertainty. To evaluate the impact of different distribution models on our results, we also consider three alternative density profiles: the Einasto profile~\cite{Navarro:2008kc}, the Isothermal profile~\cite{Bahcall:1980fb} and the Burkert profile~\cite{Burkert:1995yz}:
\begin{align}
\rho_{\rm Ein}&=\rho_{\rm s}\lt[{-\f{2}{\alpha_{\rm{Ein}}}\lt(\f{r}{r_{\rm s}}-1\rt)}\rt]
,\\
\rho_{\rm ISO}&=\f{\rho_{\rm s}}{1+(r/r_{\rm s})^2}
,\\
\rho_{\rm Burkert}&=\f{\rho_{\rm s}}{\lt[1+(r/r_{\rm s})^2\rt](1+(r/r_{\rm s})}.
\end{align}
The characteristic density $\rho_{\rm s}$, characteristic radius $r_{\rm s}$ and $\alpha_{\rm Ein}$ of these profiles are shown in Tab. \ref{tab:profile}. Further analysis of the effects induced by these different profiles will be presented and discussed in Sec.~\ref{sec:res}.
\begin{table}[!htb]
\renewcommand\arraystretch{1.3}
\centering
\begin{tabular}{cccc}
\toprule
 & $r_{\rm s}$ & $\rho_{\rm s}$ & $\alpha_{\rm Ein}$  \\
 & (kpc) & (GeV cm$^{-3}$) & \\
\midrule
NFW & 20.00 & 0.26 & $\cdots$ \\

Einasto & 20.00 & 0.06 & 0.17 \\

Isothermal & 5.00  & 1.16 & $\cdots$ \\

Burkert & 12.67  & 0.71 & $\cdots$ \\

\bottomrule
\end{tabular}
\caption{The parameters of DM density distribution profiles \cite{Bertone:2008xr,Cirelli:2010xx}.}\label{tab:profile}
\end{table}

With a local density $\rho_0 = 0.3\,\rm{GeV\,cm^{-3}}$~\cite{Bertone:2008xr,Cirelli:2010xx}, we consider sub-GeV DM particles annihilating into various final states, including $e^+e^-$, $\mu^+\mu^-$, 4 electrons/positrons, 4 muons and $\pi^+\pi^-$. Final states involving muons or pions subsequently decay, producing secondary electrons/positrons and the photons that contribute to the observed CR and MeV photon spectrum. 
These processes have been studied in previous work and can be calculated through ``PPPC4''~\cite{Cirelli:2010xx}, where the authors simulated the full spectrum using the Pythia package~\cite{Sjostrand:2007gs}.
For the specific case of electrons produced from the pion annihilation channel, we employ the HAZMA code~\cite{Coogan:2019qpu}, a publicly available tool specialized in modeling sub-GeV DM annihilation and decay.

Unlike photons, the propagation of charged CRs is significantly influenced by the local environment, including magnetic fields, interstellar gas, and turbulence. These factors can alter the energy distributions of charged particles, leading to effects such as diffusion, reacceleration, and energy losses. To calculate the propagation of these secondary electrons and positrons, we use the public code LikeDM~\cite{Huang:2016pxg}, which is based on a Green function approach derived from numerical tables provided by GALPROP~\cite{Strong:1998pw}. LikeDM has been validated to produce results consistent with GALPROP while offering significantly improved computational efficiency. The propagation parameters we used are listed in Tab. \ref{tab:propagation}, including the diffusion coefficient $D_0$, the Alfvenic speed $v_{\rm A}$ (which characterizes the reacceleration effect) and the half-height of the propagation cynlinder $z_{\rm h}$. To account for uncertainties in CR propagation, we adopt a set of propagation parameters, which allows us to evaluate the robustness of our results. These parameters are determined by fitting to the Boron-to-Carbon ratio data and the Fermi diffuse $\gamma$-ray emission. Further details can be found in~\cite{Huang:2016pxg,Fermi-LAT:2012edv,Fermi-LAT:2012pls}.

\begin{table}[htb]
\renewcommand\arraystretch{1.3}
\centering
\begin{tabular}
{ccccc}
\toprule
 & $D_0$ & $z_{\rm h}$ & $v_{\rm A}$ & $\dt$ \\
 & ($10^{28}$ ${\rm cm}^2\,{\rm s}^{-1}$) & (kpc) & (km s$^{-1}$) & \\
\midrule
Prop. 1 & 2.7  & 2 & 35.0 & 0.33 \\

Prop. 2 & 5.3  & 4 & 33.5 & 0.33 \\

Prop. 3 & 7.1  & 6 & 31.1 & 0.33 \\

Prop. 4 & 8.3 & 8 & 29.5 & 0.33 \\

Prop. 5 & 9.4 & 10 & 28.6 & 0.33 \\

Prop. 6 & 10.0 & 15 & 26.3 & 0.33 \\
\bottomrule
\end{tabular}
\caption{Propagation parameters with $z_{\rm h}$ varying from $2\,{\rm kpc}$ to $15\,{\rm kpc}$ 
Suppose a homogeneous spatial diffusion coefficient $D_{xx}=D_0\beta(E/4\,{\rm GeV})^\delta$, where $\beta$ is the Lorentz factor, $D_0$ is a coefficient, and $\delta=0.33$ reflects the Kolmogrov-type interstellar medium turbulence. The Alfvenic speed $v_{\rm A}$ characterizes the reacceleration effect. } \label{tab:propagation}
\end{table}

\section{CR Background} \label{sec:back}

To detect the signal of DM annihilation products in CR, it is essential to accurately model the astrophysical CR background. This background consists of primary electrons from sources such as supernova remnants and pulsars, as well as secondary electrons and positrons produced by inelastic collisions between CR nuclei and the interstellar medium.

The injection spectrum of primary electrons is modeled using a three-segment broken power-law with an exponential cutoff. The spectrum is given by the following equation~\cite{Zu:2021odn,Yuan:2013eja,Yuan:2013eba}:
\begin{align}\label{e_in}
\Phi_{e^{-}}&=A_{e^{-}}E^{-\nu_1^{e^{-}}}[1+(E/E^{e^{-}}_{\rm br1})^3]^{(\nu_1^{e^{-}}-\nu_2^{e^{-}})/3}\nonumber\\
&\quad\times[1+(E/E^{e^{-}}_{\rm br2})^3]^{(\nu_2^{e^{-}}-\nu_3^{e^{-}})/3}\exp(-E/E^{e^{-}}_{\rm c}).
\end{align}
where $A_{e^-}$ is the normalized factor and $\nu_{1,2,3}^{e^-}$ are the power-law indices for the three segments. The first spectral break ($E^{e^-}_{\rm br1}$), at a few GeV, accounts for low-energy data, while the second break ($E^{e^-}_{\rm br2}$), at several tens of GeV, captures the observed spectral hardening. The exponential cutoff ($E^{e^-}_{\rm c}$) is introduced to fit high-energy data, though its effect on the energy range considered here is minimal~\cite{Feng:2013zca,Yuan:2013eba,Li:2014csu}. 

Secondary electrons and positrons are produced by inelastic collisions between CR nuclei and the interstellar medium. The spectrum of secondary positron resulting from $pp$ collisions is calculated by GALPROP. During the fitting process, the flux is scaled by a constant factor to accommodate potential uncertainties in the theoretical predictions~\cite{Lin:2014vja}.

Additionally, a pulsar-like component is incorporated into the model. The injection spectrum of electrons and positrons from pulsars follows an exponential cutoff power-law~\cite{Zu:2021odn,Yuan:2013eja,Yuan:2013eba}:
\begin{equation}
\Phi_{\rm{psr}} = A_{\rm{psr}} E^{-\nu^{\rm{psr}}} \exp(-E/E^{\rm{psr}}_{\rm c}),
\end{equation}
where $E^{\rm psr}_{\rm c}$ is the cutoff for the pulsar component and $\nu^{\rm psr}, A_{\rm psr}$ represent the power index and normalized parameter. The spatial distribution of pulsars is assumed to follow that of primary CR sources~\cite{Strong:1998pw}. After the astrophysical background is well-modeled, the residual flux (after subtracting the background) can be analyzed for potential signals of DM annihilation.


\section{Results from CRs} \label{sec:res}

With the injection spectra of different source components, we can calculate the corresponding propagated spectra and compare them with observational data. Assuming the DM annihilates to $e^+e^-$, $\mu^+\mu^-$ or $\pi^+\pi^-$ within the Milky Way halo, we employ a maximum likelihood fitting approach to search for the DM component. The CR data we used includes the AMS-02 positron fraction and the total electron plus positron flux~\cite{AMS:2013fma,AMS:2014gdf}. We calculate the $\chi^2$ for each DM parameter and find the minimum $\chi^2$ as the best-fit $\chi_0^2$. Then, by setting $\Delta\chi^2 =\chi^2-\chi_0^2>2.71$, we derive the $2\sigma$ upper limits of the DM annihilation cross section. 

\begin{figure}
\centering 
\subfigure[]{\includegraphics[width=1\linewidth]{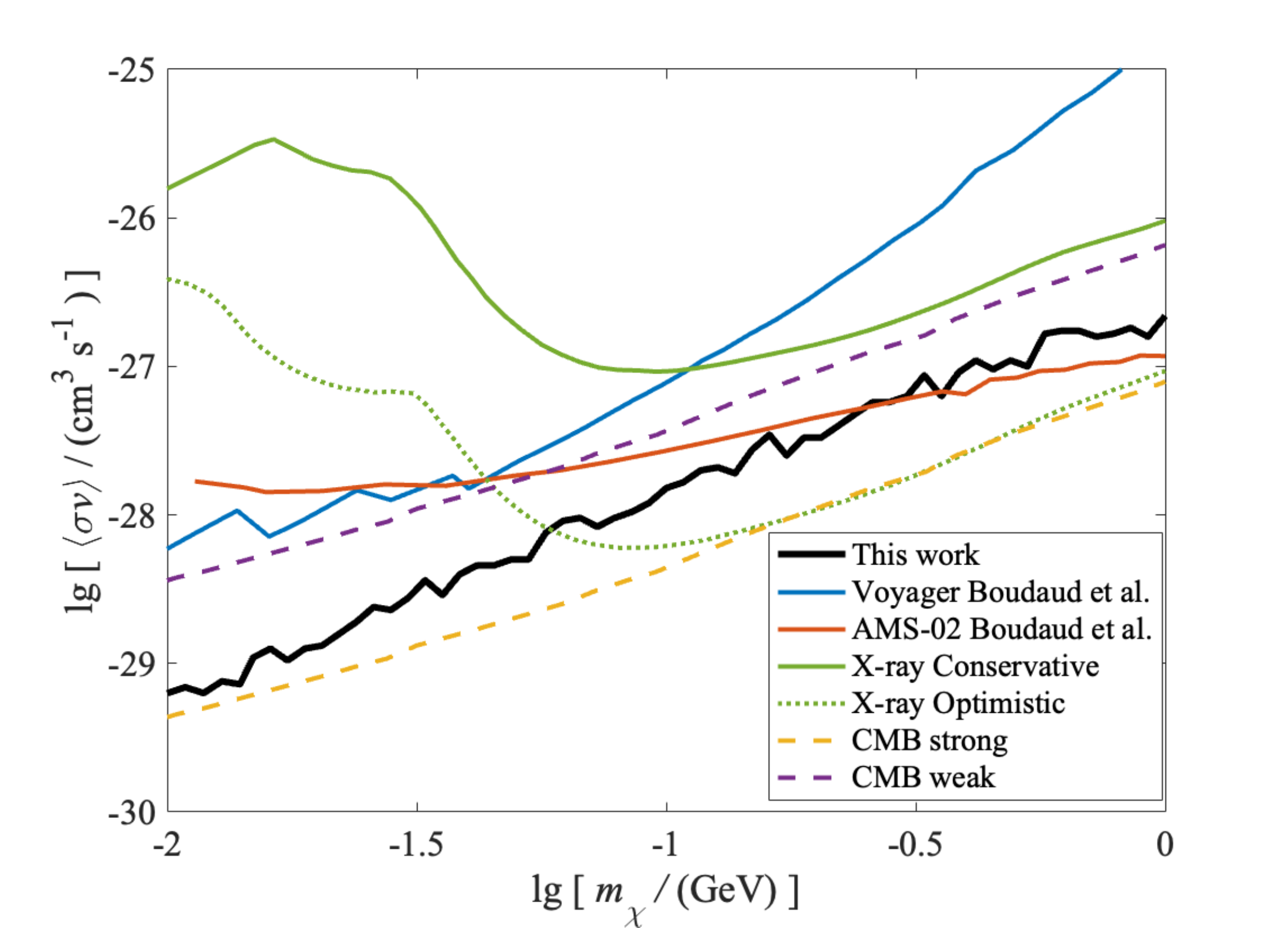}\label{fig:Electron}} \\
\subfigure[]{\includegraphics[width=1\linewidth]{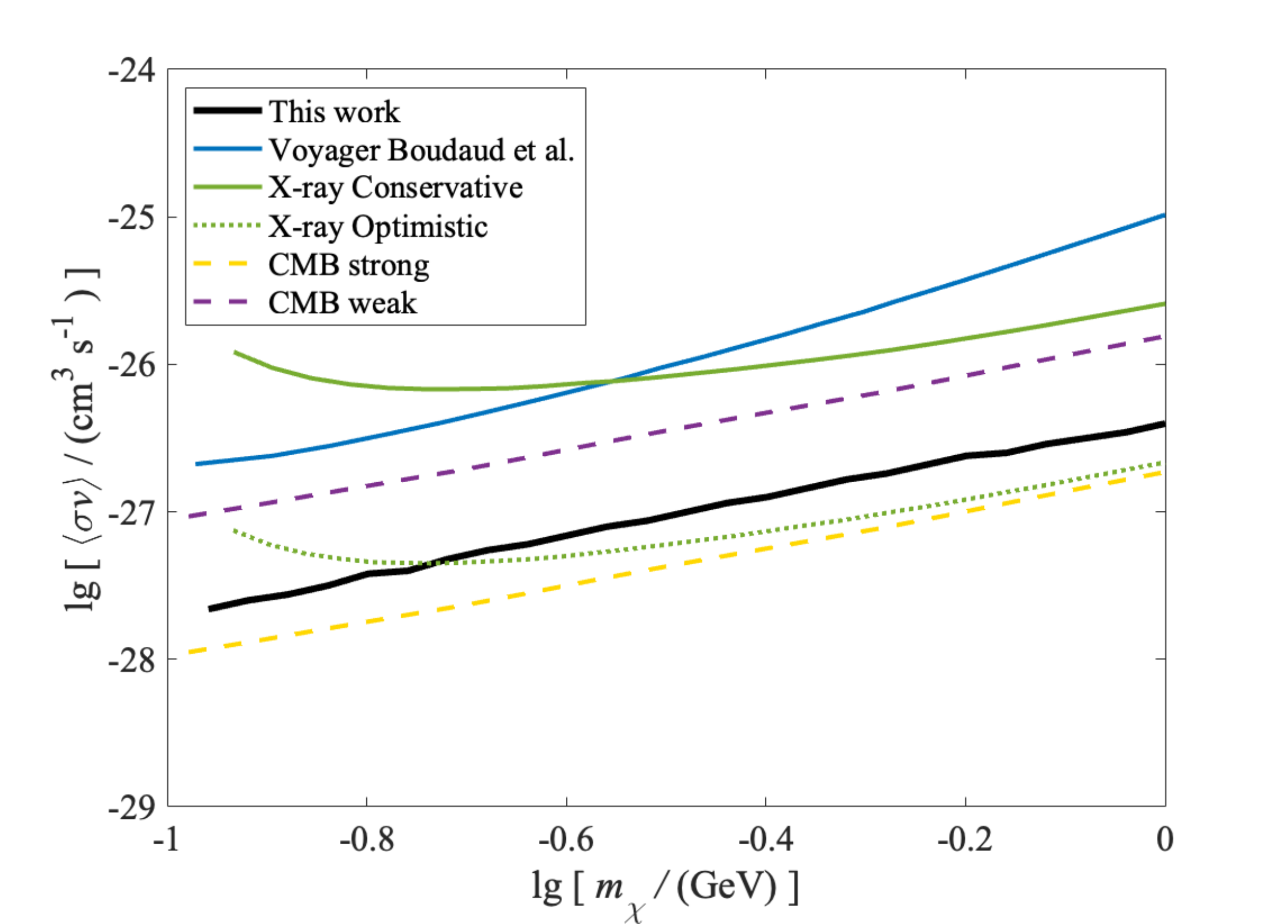}\label{fig:Muon}}
\caption{Our constrains of DM annihilation cross section $\langle\sigma v\rangle$ (black solid lines), compared with existing results from: Voyager data (blue solid lines) and AMS-02 data (red solid line) by M.~Boudaud~et~al.~\cite{Boudaud:2016mos}, X-ray observations (green solid and dotted lines) by M.~Cirelli~et~al.~\cite{Cirelli:2020bpc,Cirelli:2023tnx}, and CMB analyses (yellow and purple dashed lines)~\cite{Slatyer:2015jla,Leane:2018kjk}. Figure~\ref{fig:Electron} shows the results for $e^+e^-$, and Fig.~\ref{fig:Muon} presents those for $\mu^+\mu^-$.} \label{fig:results}
\end{figure}

In the following works, we use Prop. 2 from Tab.~\ref{tab:propagation} as an example to illustrate the constraints. If DM annihilates into $e^+e^-$ or $\mu^{+}\mu^-$, we present the 95\% upper limits on the annihilation cross-section in Fig. \ref{fig:results}. For DM masses ranging from 100 MeV to 1 GeV, our derived limits on the annihilation cross-section lie between $10^{-28}$ and $10^{-27}\,$cm$^3\,$s$^{-1}$, which are comparable to the results from CMB~\cite{Slatyer:2015jla,Leane:2018kjk} and are more stringent than those derived from Voyager data~\cite{Boudaud:2016mos} for more than an order of magnitude. 
Also, we compare our constraints with the X-ray bounds from Refs.~\cite{Cirelli:2020bpc,Cirelli:2023tnx}, which provide both conservative (without including astrophysical backgrounds) and optimistic (with backgrounds) limits. Our results exceed the conservative limits from X-ray observations (green solid line). However, when compared to the optimistic bounds (green dotted line), our constraints are competitive and even stronger in the low-mass region, whereas the X-ray analysis provides stronger constraints at higher masses. This is because, at lower masses, the X-ray signal from inverse-Compton scattering falls below the detectable energy range, weakening their limits~\cite{Cirelli:2020bpc,Cirelli:2023tnx}.
Moreover, as shown in Fig.~\ref{fig:Electron}, our results (black line) provide stronger constraints on the DM annihilation cross section for DM masses below approximately 300 MeV, surpassing results from Ref.~\cite{Boudaud:2016mos} (red line) by about an order of magnitude in the several-tens-of-MeV region.
This is because they only considered the secondary electron background and neglected the contribution from primary electrons originating from pulsars or supernova remnants.
Furthermore, we adopt a lower local DM density of $\rho_0 = 0.3\,\rm{GeV\,cm^{-3}}$~\cite{Bertone:2008xr}, compared to their value of $0.4\,\rm{GeV\,cm^{-3}}$. 
In addition, we also employ updated CR propagation parameters derived from the Boron-to-Carbon ratio data and the Fermi diffuse $\gamma$-ray emission~\cite{Huang:2016pxg,Fermi-LAT:2012edv,Fermi-LAT:2012pls}, which are much more reliable. The annihilation of DM into charged particles and photons is constrained by observations of the CMB, which influence the ionization fraction and modify CMB anisotropies. Previous studies indicate that the limit is set as $f_{\text{eff}} \langle\sigma v\rangle/2m_\chi < 4.1 \times 10^{-28} \, \text{cm}^3\,\text{s}^{-1}\,\text{GeV}^{-1}$, where $ f_{\text{eff}} $ is the effective efficiency factor. In this analysis, we use $ f_{\text{eff}} = 0.08 $ for a conservative limit and $ f_{\text{eff}} = 1 $ for a progressive limit, leading to the constraints $\langle\sigma v\rangle_{\bar{f}{f}}/{(2 m_\chi)} \leq 5.1 \times 10^{-27} \, \text{cm}^3\,\text{s}^{-1}\,\text{GeV}^{-1}$ and $\langle\sigma v\rangle_{\bar{f}{f}}/{(2 m_\chi)} \leq 4.1 \times 10^{-28} \, \text{cm}^3 \,\text{s}^{-1}\,\text{GeV}^{-1}$ \cite{Slatyer:2015jla,Leane:2018kjk}.

\begin{figure}
\centering 
\subfigure[]{\includegraphics[width=1\linewidth]{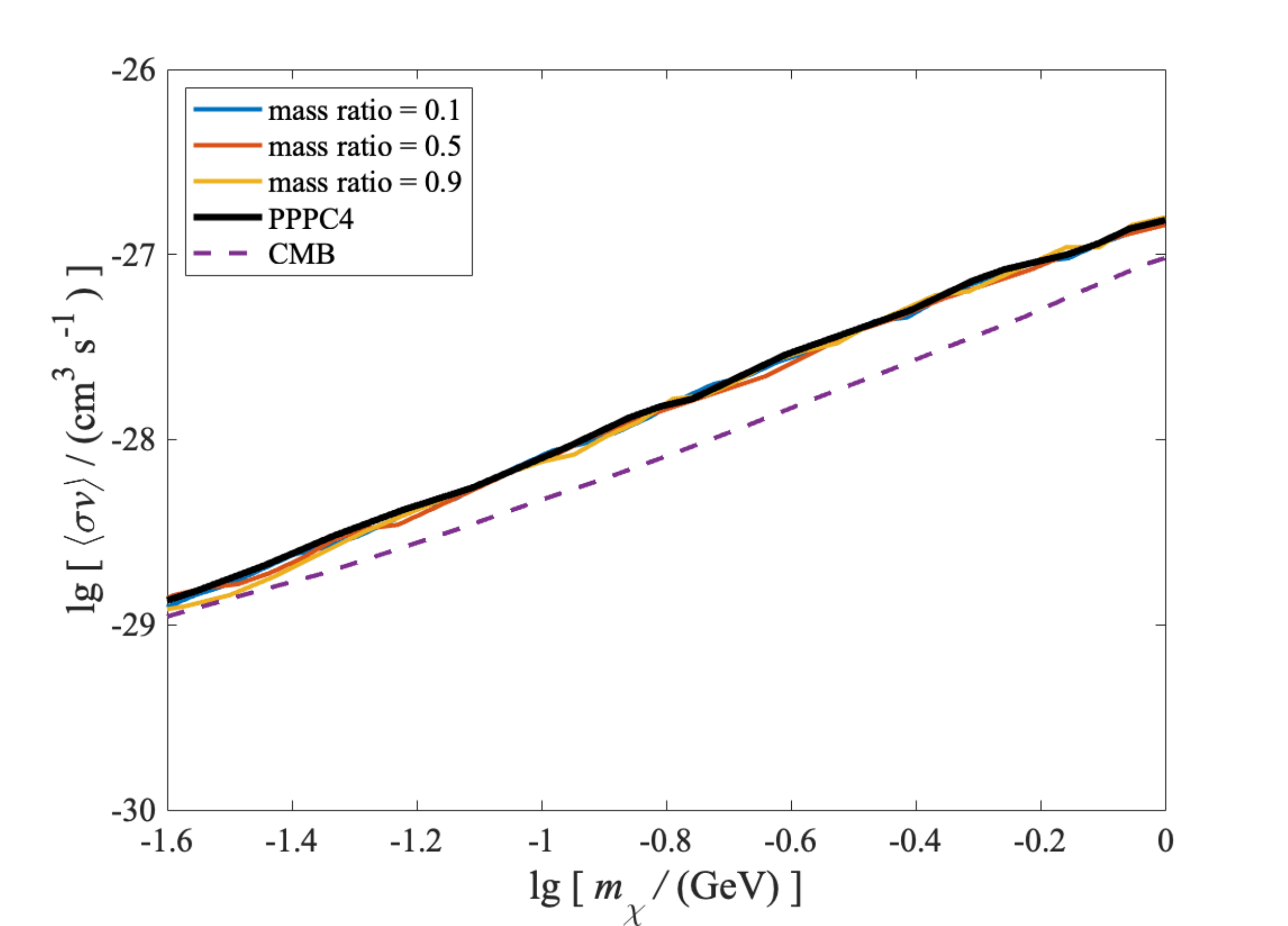}\label{fig:4 electrons}} \\
\subfigure[]{\includegraphics[width=1\linewidth]{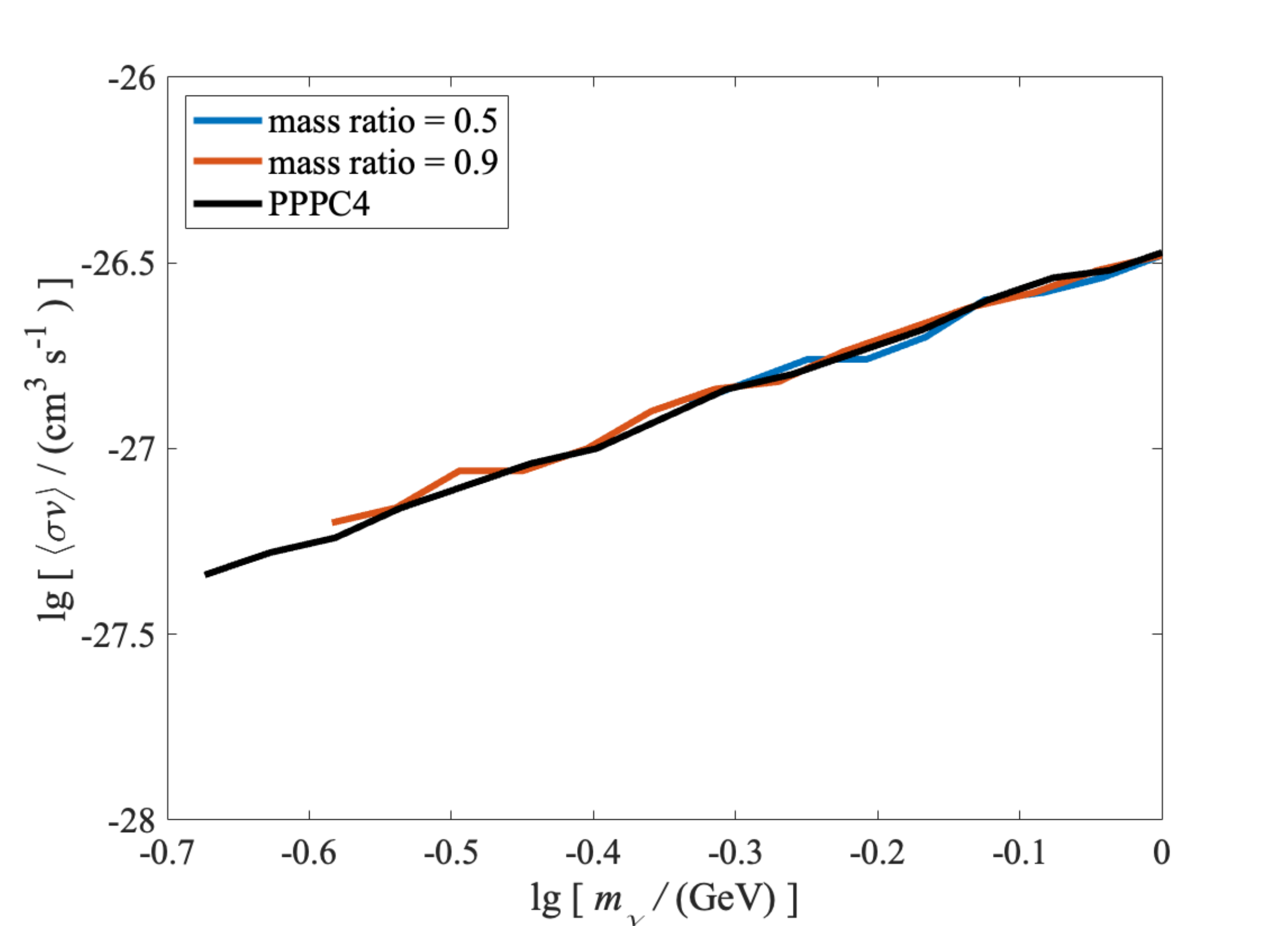}\label{fig:4 muons}}
\caption{Our constrains of DM annihilation cross section $\langle\sigma v\rangle$ (black solid lines). The results of 4 electrons are shown in Fig. \ref{fig:4 electrons} (compared with results from CMB analyses (purple dashed line)~\cite{Slatyer:2015jla,Leane:2018kjk}), along with those of 4 muons in Fig. \ref{fig:4 muons}.}
\label{fig:results2}
\end{figure}

\begin{figure}[htb]
\centering
\includegraphics[width=\linewidth]{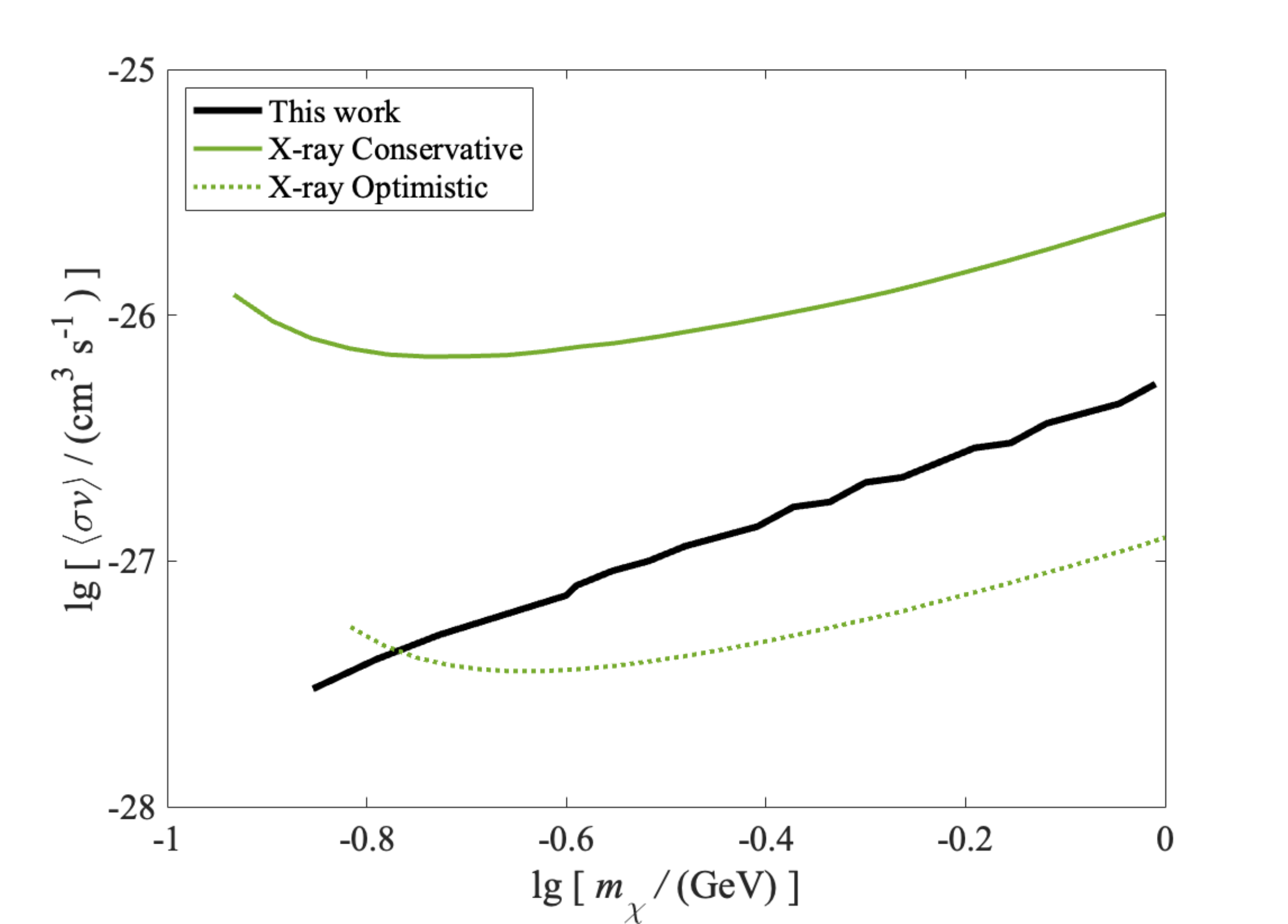}
\caption{Our constrains of DM annihilation cross section $\langle\sigma v\rangle$ for pion channel (black solid line), compared with results from X-ray observations (green solid and dotted lines) by M.~Cirelli~et~al.~\cite{Cirelli:2020bpc,Cirelli:2023tnx}.}
\label{fig:pion}
\end{figure}

Additionally, if DM annihilates into four electrons or four muons via light mediators, this study provides new constraints on these less explored annihilation channels, probing different aspects of DM interactions (see Fig.~\ref{fig:results2}). The calculations of DM particles annihilation in these two processes were conducted using PPPC4. To further investigate the impact of the intermediate particle mass on the results, we examined scenarios where the intermediate particle masses are set to 0.1, 0.5, and 0.9 times the DM mass. The results were compared with those from PPPC4, and they were found to be nearly identical. This consistency suggests that the intermediate particle mass, within the considered range, has a negligible effect on the final outcomes. The pion results are also shown in Fig.~\ref{fig:pion}. Similar to electron and muon channels, our results are comparable to the optimistic X-ray bounds at low energies. Also, even though their constraints become weaker at higher energies, they remain stronger than the conservative X-ray limits.

\begin{figure}
\centering 
\subfigure[]{\includegraphics[width=1\linewidth]{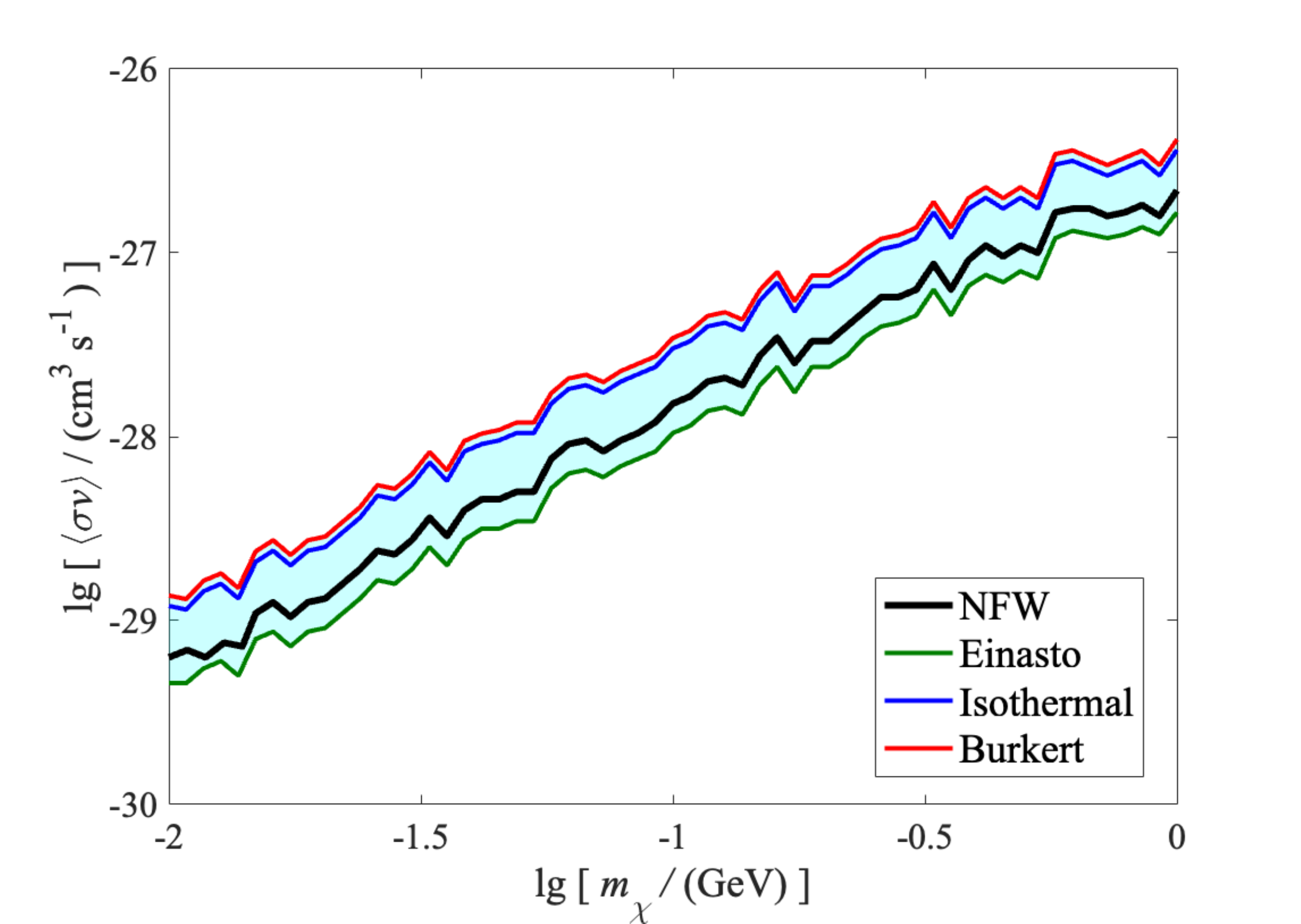}\label{fig:DMprofile}} \\
\subfigure[]{\includegraphics[width=1\linewidth]{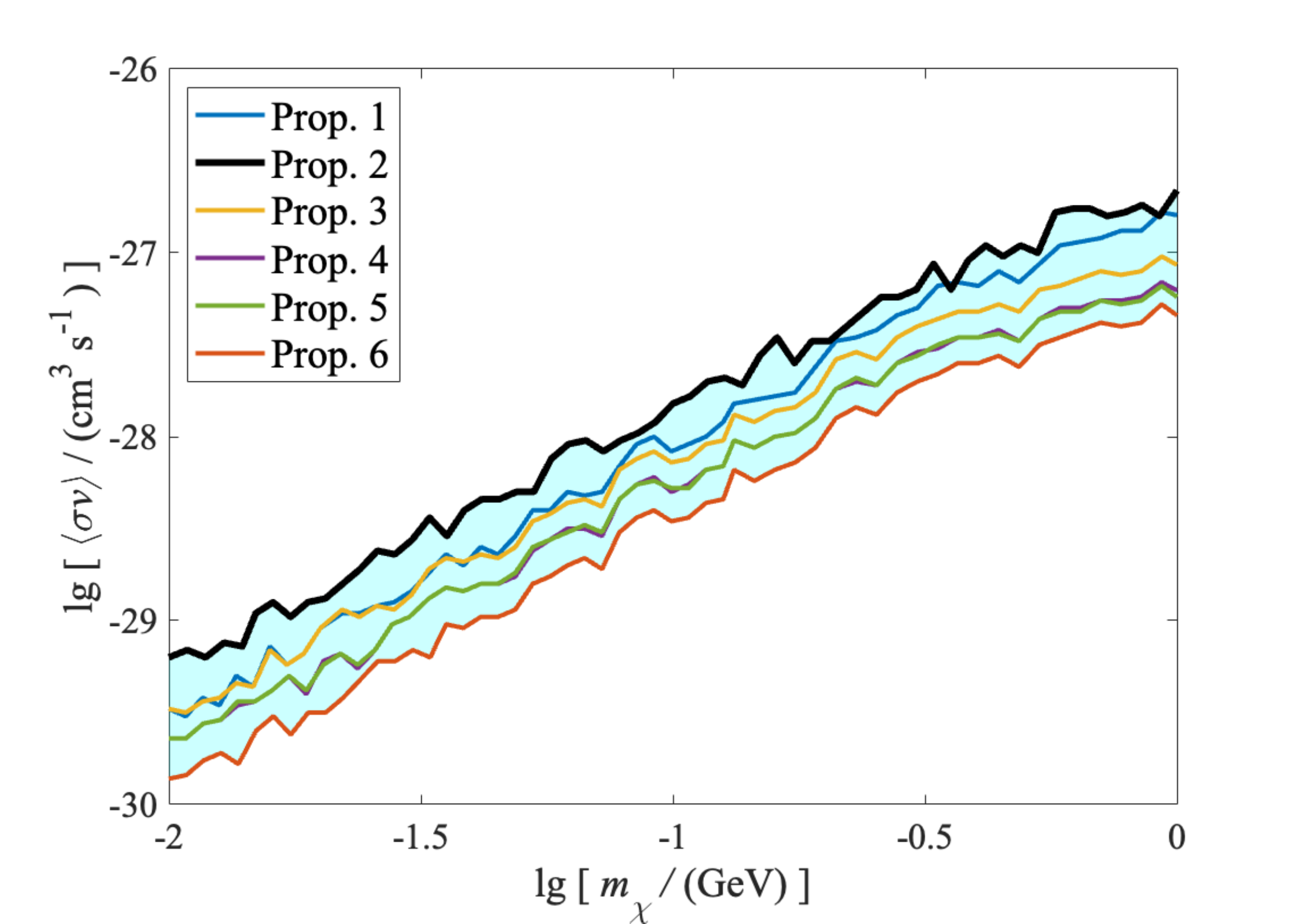}\label{fig:prop}}
\caption{Impact of parameter uncertainties on the limits of DM annihilation cross section $\langle\sigma v\rangle$ in $e^+e^-$ channel. Fig. \ref{fig:DMprofile}: Different DM density profile models. Fig. \ref{fig:prop}: Different propagation parameters.}
\label{fig:discuss}
\end{figure}

To assess the uncertainties in our results, we analyze the impact of both DM density distribution profile models and the CR propagation parameters. Specifically, we adopt four DM profiles (NFW, Einasto, Isothermal, and Burkert profiles) summarized in Tab.~\ref{tab:profile}, along with six distinct propagation models (Props. 1–6) listed in Tab.~\ref{tab:propagation}. Taking the $e^+ e^-$ channel as an example, the results are shown in Fig.~\ref{fig:discuss}. In Fig.~\ref{fig:DMprofile}, the black, green, blue, and red lines correspond to the NFW, Einasto, Isothermal, and Burkert profiles, respectively.
It is clear that the Burkert profile provides the most conservative constraints, whereas the Einasto profile yields the strongest limits. In Fig.~\ref{fig:prop}, we find that Prop.~2, which was used to derive the constraints presented earlier, provides the most conservative results among the set. Nevertheless, all profiles are in good agreement with each other within one order of magnitude.

\section{Results from photons} \label{sec:gamma}

When the DM annihilates to some charged particles, final state radiation of photons is inevitably produced. These photons provide complementary information to the study of DM annihilation. Unlike charged particles, photons are not affected by interstellar magnetic fields and propagate directly, offering a clearer signature of the annihilation process. By combining photon signals with cosmic ray observations, a multi-messenger approach can be employed to enhance the detection capability and scientific understanding of DM annihilation. 

In this section, we investigate photon emissions resulting from DM annihilation in the Galactic Center (GC), complementing our study of DM annihilation processes in the broader Milky Way. The GC is a region of particular interest due to its high predicted DM density, which makes it a prime target for detecting potential DM annihilation photon signals. The photons from the final state radiation primarily occurs at the sub-GeV scale. While no high-sensitivity telescopes currently operate in this energy range, several proposed missions, such as the All-sky Medium Energy Gamma-ray Observatory (AMEGO)~\cite{AMEGO:2019gny} and Very Large Area gamma-ray Space Telescope(VLAST)~\cite{fanyizhong}, are expected to be launched in the near future, significantly enhancing our understanding of sub-GeV astrophysical phenomena. 

\begin{figure}[htb]
\centering
\includegraphics[width=\linewidth]{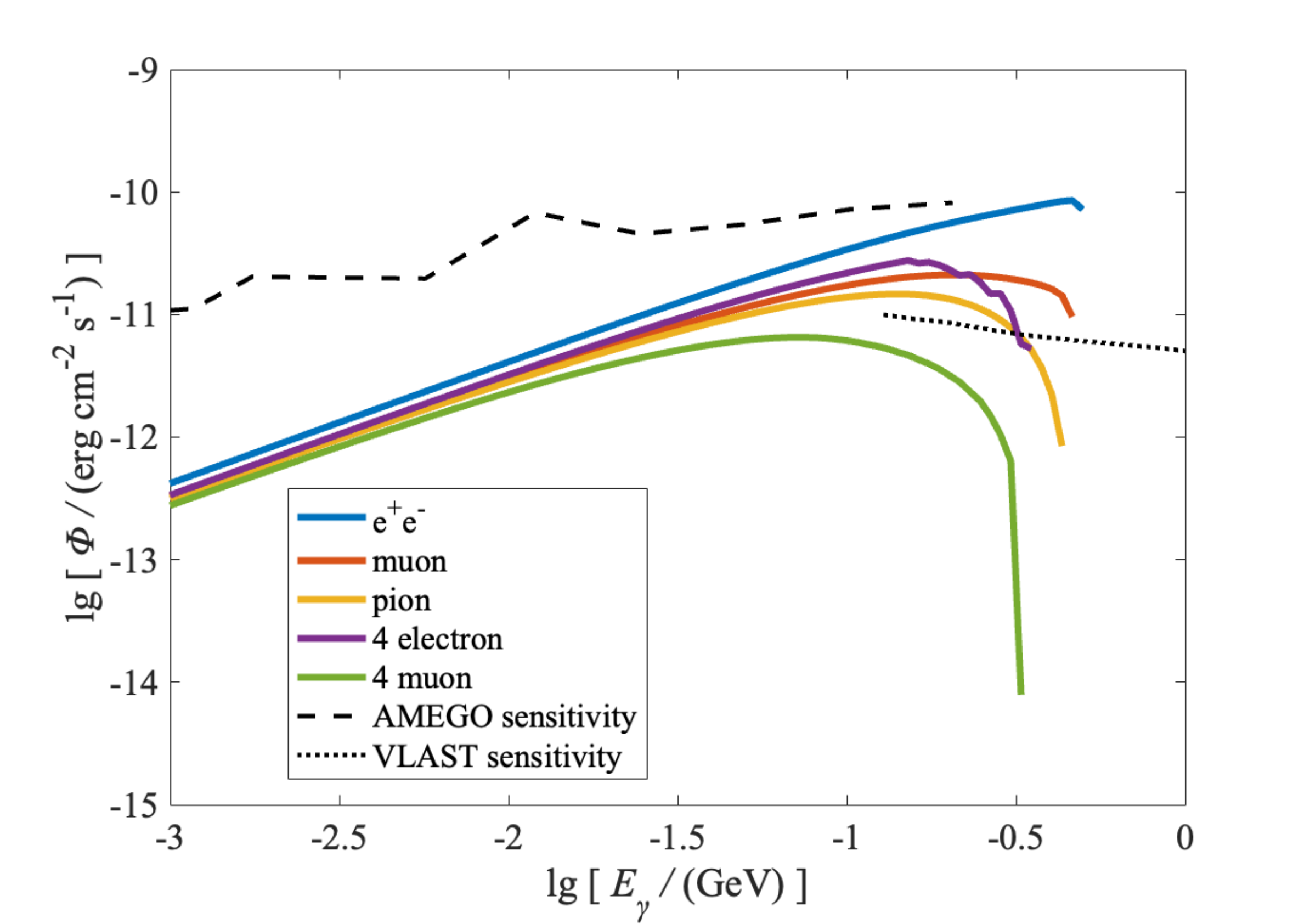}
\caption{The energy flux of gamma-ray photons resulting from the final state radiation of DM in the Galactic halo across different annihilation channels is depicted. The black dashed line represents the sensitivity of AMEGO~\cite{AMEGO:2019gny,Negro:2021urm}, while the dotted line corresponds to VLAST~\cite{fanyizhong}.}
\label{fig:gamma}
\end{figure}

In this analysis, we adopt the same DM density profile as used in the CR study. This profile yields a J-factor of $2.503\times10^{22}~\rm{GeV^2~cm^{-5}}$ within a circular region with a radius of 5 degrees~\cite{Huang:2016pxg}, which quantifies the line-of-sight integral of the squared DM density and is critical for calculating the expected annihilation flux. We focus on a DM mass of 500 MeV and compute the final state radiation photon fluxes for various annihilation channels. The results are presented as solid lines in the Fig.~\ref{fig:gamma}, with different colors corresponding to different final states. 

The predicted photon fluxes are compared with the sensitivity curves of upcoming gamma-ray telescopes including the AMEGO and VLAST, represented as dashed and dotted lines in the figure\footnote{The extended sources sensitivities of AMEGO is calculated by the point sources sensitivities follow the method in~\cite{Negro:2021urm}).}. The expected photon flux from DM annihilation has been found to reach the sensitivity thresholds of VLAST. The results of 4 muons channel do not reach the sensitivities of AMEGO and VLAST, which implies that we may not get stronger constrains in this specific channel with future gamma-ray telescopes. However, in others channels, the results demonstrate that future telescopes will play a crucial role in constraining the properties of sub GeV DM and could potentially uncover definitive evidence of DM annihilation in the Milky Way.
 
 \section{Conclusion} \label{sec:con}
 
The search for DM annihilation signals through CRs and future MeV gamma-ray telescopes represents a powerful and complementary approach to unraveling the mysteries of DM. By analyzing CR data from experiments like AMS-02, we have constrained the sub-GeV DM annihilation cross-section in the range of $10^{-28}$ to $10^{-27}$ cm$^3\,$s$^{-1}$, corresponding to DM masses ranging from 100 MeV to 1 GeV. However, the limitations of CR detection due to the uncertainty of the astrophysical backgrounds and propagation process, highlight the need for alternative detection methods.

Future MeV gamma-ray telescopes, such as AMEGO and VLAST, offer a promising avenue to overcome these challenges. These instruments are designed to bridge the "MeV gap," providing unprecedented sensitivity to photon signals from DM annihilation. Our analysis demonstrates that the expected photon fluxes from DM-rich regions, such as the GC, could reach the detection thresholds of these telescopes.

The combination of CR and MeV gamma-ray observations enables a multi-messenger approach to DM detection. While CRs provide insights into charged particle products of DM annihilation, MeV photons offer a direct and complementary probe of the annihilation process, free from the confounding effects of interstellar magnetic fields. Together, these methods enhance our ability to disentangle DM signals from astrophysical backgrounds and constrain DM properties with greater precision.

\vspace{.5cm}
\noindent {\bf Acknowledgements} We thank Qiang Yuan for kindly providing the relevant code. This work is supported by the National Key R\&D Program of China (Grant No. 2022YFF0503304), the National Natural Science Foundation of China (Grant Nos. 12373002, 12220101003, and 11773075), and the Youth Innovation Promotion Association of Chinese Academy of Sciences (Grant No. 2016288). LZ is supported by the NAWA Ulam fellowship (No. BPN/ULM/2023/1/00107/U/00001) and the National Science Centre, Poland (research grant No. 2021/42/E/ST2/00031).

\vspace{.5cm}
\noindent {\bf Data Availability Statement} No Data associated in the manuscript.

\end{document}